\documentclass[reprint,amsmath,amssymb,graphicx,longbibliography,showpacs]{revtex4-1}

\usepackage{amssymb}

\usepackage{amsfonts}
\usepackage{latexsym}

\usepackage{cancel}
\usepackage{amsmath}
\usepackage{braket}
\usepackage{graphicx}

\usepackage{psfrag,pstricks}
\usepackage[breaklinks=true,colorlinks=true,linkcolor=blue,urlcolor=blue,citecolor=blue]{hyperref}

\begin{document}

\title{ Generating quantum discord between two distant Bose-Einstein condensates with Bell-like detection }
\author{M. Eghbali-Arani$^{1, 2}$}
\author{H. Yavari$^{2}$}
\author{M. A. Shahzamanian$^{2}$}
\author{V. Giovannetti$^{1}$}
\author{Sh. Barzanjeh$^{3}$}
\affiliation{
$^1$ NEST, Scuola Normale Superiore and Istituto Nanoscienze-CNR, I-56127 Pisa, Italy\\
$^2$ Department of Physics, Faculty of Science, University of Isfahan, Hezar Jerib, 81746-73441, Isfahan, Iran \\
$^3$ Institute for Quantum Information, RWTH Aachen University, 52056 Aachen, Germany}
\begin{abstract}
We propose a technique that enables the creation of quantum discord between two distant nodes, each containing a cavity consist of the Bose-Einstein condensate, by applying a non-ideal Bell-like detection on the output modes of optical cavities. We find the covariance matrix of the system after the non-ideal Bell-like detection, showing explicitly that one enables manipulation of the quantum correlations, and particularly quantum discord, between remote Bose-Einstein condensates. We also find that the non-ideal Bell-like detection can create entanglement between distant Bose-Einstein condensates at the two remote sites. 
\end{abstract}

\maketitle

\section{Introduction}
Nonlocal entanglement is one of the most fundamental features
of quantum mechanics and it has variety of 
applications in quantum information processing~\cite{Nielsen,Wilde,Cerf, Braunstein}, such as secret key distribution~\cite{Ekert},
quantum teleportation~\cite{Bennett1,Braunstein2, Bennett} and superdense coding~\cite{
 Ban, Braunstein3, Ban2}. One possible way to establish nonlocal entanglement is to
perform a projection measurement on the state of two particles onto
an entangled state. This measurement does not
necessarily require a direct interaction between the two
particles. If each of the particles is entangled with
one other partner particle, an appropriate measurement~(such a Bell-state measurement) on the partner particle
will collapse the state of the remaining two
particles onto an entangled state. This Bell projection, which establishes nonlocal correlations
between 
distant systems that have never interacted is a key ingredient in quantum computation and
communication protocols such as teleportation~\cite{Bennett1} and
entanglement swapping~\cite{Zukowski, Loock, Bou, Cerf}. 

Entanglement swapping is the central principle used
in quantum repeaters~\cite{Zoller}, whose main aim is to overcome photon loss in long-range quantum
communication. Entanglement swapping can be used as a realistic protocol to generate entanglement between quantum memories at distant nodes in hybrid quantum repeaters; it has been implemented in several discrete~\cite{Bose} and continuous variable~(CV)~\cite{Tan, Ban4, Megidish, Jia, Furo,Abdi, hamm, pirandola} systems. 

Entanglement swapping, however, is able to generate the other types of quantum correlations such as quantum discord between distant sites. Quantum discord as a measure of the quantumness of correlations is important because
 entanglement does not describe all the non-classical properties of quantum correlations. Quantum discord has been proposed as a measure of all quantum correlations  including, but not restricted to, entanglement~\cite{Zurek, Zurek2}. It is known that the quantum discord can offer a new model of quantum information processing which requires very little or no
entanglement~\cite{bor}. Therefore, the quantum discord proposes that the separable or weakly entangled states could be used as a powerful tools for quantum information processing~(such as quantum remote state preparation) as
they are much easier to prepare and control even in dissipative environment.

Motivated by the above mentioned studies, in this paper we propose a scheme for generating quantum correlation between two distant nodes by applying a non-ideal Bell-like detection. In such scheme each node
contains a Bose-Einstein condensate~(BEC)~\cite{Dalafi1, Nagy} trapped inside an optical cavity~[see Fig.~\ref{f1}.(a)]. The output of the optical cavity of each node (which shows robustness against thermal noise) is sent to an intermediate site where a Bell-like detection is performed on these optical pairs~[see Fig.~
\ref{f1}.(b)]. We show that the immediate consequence of this Bell-like detection is the creation of quantum discord between two distant BECs.  Moreover, our system can be
realized with state-of-the-art technology, and is suited
to such potential applications as quantum remote state preparation and quantum state teleportation between distant BECs.
  
This paper is organized as follows. In Sec. \ref{formulation} we provide the theoretical description of the cavity containing a BEC.  In Sec. \ref{protocol} we introduce our proposal for producing quantum discord between two distant nodes by using a non-ideal Bell-like detection. We find the covariance matrix after Bell measurement and then we explicitly show that the Bell-like detection can produce quantum discord between remote BECs. In Sec. \ref{entanglement} we discuss
the generation of bipartite entanglement between two distant BECs. Finally, our conclusions are given in Sec. \ref{conclusion}.

\section{Formulation and Theoretical Description of The System at Each Node}\label{formulation}

\begin{figure}[ht]
\centering
\includegraphics[width=3in]{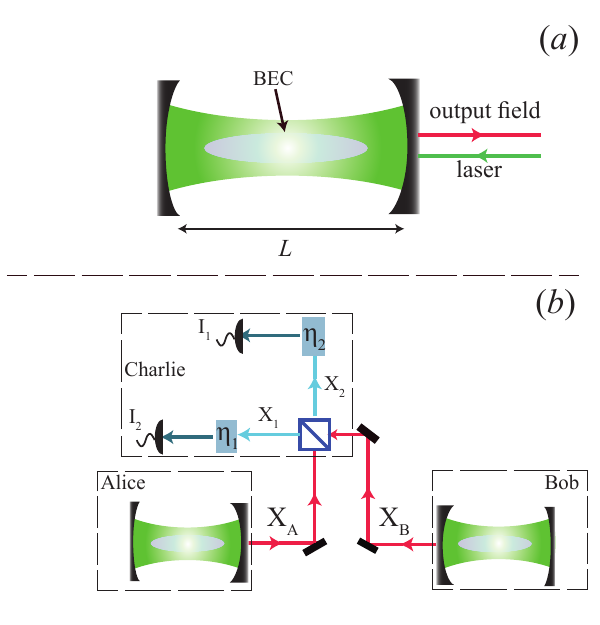}
\caption{(Color online)~(a)~Scheme of a trapped ultracold atoms inside the cavity where the BEC atoms are coupled with a single mode of the optical cavity. We also assume the optical cavity is driven by an intense laser. (b)~A possible protocol for generating the quantum correlations between two BECs by applying a non-ideal Bell-like detection.}
\label{f1}
\end{figure}

Our main goal is to establish the quantum discord 
between two distant nodes that never interacted~(see Fig.~\ref{f1}b). In order to construct such quantum link first we need to study the dynamics of the system at each node. 

The schematic of the system at each node is depicted in Fig.~\ref{f1}(a).  We consider a cigar-shaped gas of $N$ ultracold bosonic two-level atoms with transition frequency $\omega_a$ and mass $m_a$  inside an optical cavity with length $ L $.  The optical cavity is driven by an intense laser at frequency $\omega_{\mathrm{d} }$ with amplitude $ E_{\mathrm d}= \sqrt{2 \mathcal{P} \kappa/\hbar \omega_{c}} $, where $\mathcal{P}$ is the power of the input laser and $ \kappa $ shows the damping rate of the optical cavity with the bare frequency $\omega_c$. 
In the dispersive regime where the laser pump is far
detuned from the atomic resonance ($\Delta_{a}=\omega_{\mathrm d}-\omega_{a}$), the excited
electronic state of the atoms can be adiabatically eliminated
and spontaneous emission can be neglected~\cite{Maschler}. 
Therefore, the many-body Hamiltonian in the rotating frame with respect to the laser frequency $\omega_{\mathrm d}$ takes the following form~\cite{Dalafi1}
\begin{eqnarray}\label{Hamiltonian}
\hat H&=& \hbar\Delta_{c} \hat{a}^\dagger \hat a+\mathrm i \hbar E_{\mathrm d} (\hat{a}^\dagger - \hat a)+\int_{-L/2}^{L/2} dx\, \hat{\psi}^\dagger (x) \Big[\frac{-\hbar^2}{2m_a}\frac{d^2}{dx^2}\nonumber\\&+&\hbar U_0 \mathrm{cos}^2(kx) \hat{a}^\dagger \hat a+\frac{1}{2}U_s\,\hat \psi^\dagger(x)\hat \psi(x) \Big]\hat \psi(x),
\end{eqnarray}
where $\hat a~([\hat a, \hat a^\dagger]=1)$ is the annihilation operator for the cavity mode, and $\Delta_{c}=\omega_{c}-\omega_{\mathrm d}$ is the detuning between the laser and a cavity mode. The parameter $U_{0}=g_{0}^2/\Delta_{a}$ is the height of potential barrier of the optical lattice per photon and represent the backaction on the field, $g_{0}$ is the vacuum Rabi frequency, $U_{s}=\frac{4\pi\hbar^2 a_{s}}{m_a}$, and $a_{s}$ is the two-body s-wave scattering length~\cite{Domokos, Maschler2}.

In equation~(\ref{Hamiltonian}) we assume only the first two symmetric momentum side modes with momenta $\pm 2k$ are excited by  the atom-light interaction~\cite{Nagy}. This approximation is valid in the weekly interacting regime i.e,. $ U_0 \langle a^{\dagger}a\rangle \leq 10 \omega_R $~($ \omega_R = \hbar k^2/2m_a$ is the recoil frequency of the condensate atoms and $k=\omega_\mathrm{d}/c$ is the wavenumber). However, because of the parity conservation and considering the Bogoliubov approximation, the atomic field operator can be expanded as
\begin{eqnarray}\label{psi}
\hat \psi(x)=\sqrt{\frac{N}{L}}+\sqrt{\frac{2}{L}}\cos(2kx)\hat c,
\end{eqnarray}
where $ \hat c\,([\hat c, \hat c^{\dagger}]=1)$.  

 By substituting the atomic field operator Eq.~(\ref{psi}), into the Hamiltonian~(\ref{Hamiltonian}) one obtains~\cite{Dalafi1}
\begin{eqnarray}\label{Hamiltonian2}
\hat H&=&\hbar \delta_{c}\hat {a}^\dagger \hat a+\frac{1}{2}\hbar\Omega_c (\hat {P}^2+\hat {Q}^2)\nonumber\\&+&\hbar G \hat {a}^\dagger \hat a \hat Q+\frac{1}{2}\hbar\omega_{sw}\hat {Q}^2+i\hbar E_{\mathrm d}(\hat {a}^\dagger-\hat a),
\end{eqnarray}
where we define the Bogoliubov position $\hat Q=(\hat c+\hat c^\dagger)/\sqrt{2}$ and momentum $\hat P=(\hat c-\hat c^\dagger)/i\sqrt{2}$ quadratures. However,
 $ \delta_{c}=\Delta_{c}+\frac{N U_{0}}{2}$ is the effective Stark shift detuning. $\omega_{sw}=\frac{8 \pi \hbar N}{M L {\nu}^2}$  is the s-wave scattering frequency and $\nu$ is the waist of the optical potential. $\Omega_c=4\omega_{R}+1/2 \omega_{sw}$ is the effective detuning of BEC and $G=\frac{\omega_c}{L}\sqrt{\frac{\hbar}{4\omega_R m_s}}$ is the coupling strength of the optical cavity mode with Bogoliubov mode, where $m_s=\frac{\hbar{\omega_c}^2}{L^2N{U_0}^2 \omega_R}$ is effective mass of side mode.

The quantum Langevin equations~(QLEs) for BEC and the cavity variables are obtained by adopting the dissipation-fluctuation theory~\cite{Zoller1}
\begin{eqnarray}\label{qlemain}
\hat{ \dot{P}}&=&-(\Omega_c+\omega_{sw})\hat Q -\gamma_c\hat P-G \hat a^\dagger \hat a+\sqrt{2\gamma_c}\hat Q_{in},\nonumber\\
\hat{ \dot{Q}}&=&\Omega_c \hat P-\gamma_c \hat Q+\sqrt{2\gamma_c}\hat P_{in},\\
\hat{ \dot{a}}&=&-(\mathrm i \delta_c+\kappa)\hat a +\mathrm i G\hat Q\hat a+E_{\mathrm{d}}+\sqrt{2\kappa}\hat a_{in}\nonumber,
\end{eqnarray}
where $ \gamma_c $ is the dissipation of the collective density excitations of the BEC, and $\hat a_{in}(t)$ is the cavity input noise which obeys the white-noise correlation functions~\cite{Zoller1}
\begin{eqnarray}
\langle \hat a_{in}(t) \hat a_{in}^{\dagger}(t^{\prime})\rangle & = &\delta(t-t^{\prime}),\nonumber\\
\langle \hat a_{in}^{\dagger}(t)\hat a_{in}(t^{\prime})\rangle &=&0,\label{coropt}
\end{eqnarray}
where we have set $N=[\mathrm{exp}(\hbar\omega_{c}%
/k_{B}T)-1]^{-1}\approx0$, since $\hbar\omega_{c}/k_{B}T\gg1$ at optical frequency, 
where $k_{B}$ being the Boltzmann constant, and $T$ is the temperature of the reservoir.
On the other hand, $ \hat Q_{in} $ and $\hat  P_{in} $ are the thermal noise inputs for the side mode of BEC which satisfy the Markovian correlation functions $ \langle \hat P_{in}(t)\hat P_{in}(t')\rangle=\langle \hat Q_{in}(t) \hat Q_{in}(t')\rangle=2\gamma_c(n_c+1/2)\delta(t-t')$~\cite{Dalafi2,Zhang}, where we have assumed that the BEC has been trapped and isolated from its environment so that its effective temperature $ T_c\simeq 0.1 \mu$K and $n_c=[\mathrm{exp}(\hbar\omega_{B}/k_{B}T_c)-1]^{-1}$ is the number of thermal excitations for the Bogoliubov mode which oscillates with frequency $ \omega_{B}=\sqrt{\Omega_c(\Omega_c+\omega_{sw})} $. 

One simple way to represent the dynamics of the system is to use the vector $\mathbf{u}=[\hat Q,\hat P, \hat X,\hat Y]^T$, where the quadratures of the optical field cavity are defined as $ \hat X=(\hat a+\hat a^{\dagger})/\sqrt{2} $ and $\hat Y=i(\hat a^{\dagger}-\hat a)/\sqrt{2} $. Since the cavity is driven by an intense laser then one can linearize the QLEs given in Eq.~(\ref{qlemain}) around the semiclassical mean values, i.e,. $u_j=u_{j,s}+\delta u_j(t)$, where $ u_{j,s} $ are the classical mean values and $\delta u_j(t)$ are zero-mean fluctuation operators. The
mean values are obtained by setting the time derivatives to zero,
resulting in
\begin{eqnarray}
Q_{s}&=&-\frac{G\alpha_s^2}{\Omega_c+\omega_{sw}+\frac{\gamma_{c}^2}{\Omega_c}}, \\ 
P_{s}&=&\frac{\gamma_{c}}{\Omega_c}Q_s,\nonumber\\
\alpha_s &=&\frac{E_\mathrm{d}}{\sqrt{\Delta^2+\kappa^2}},\nonumber
\end{eqnarray}
where $\Delta=\delta_{c}+G Q_{s}$ is the effective detuning and $ \alpha_s=\alpha_s^*$ is the mean field value of the optical cavity mode.

The dynamics of the quantum fluctuations, $\delta \textbf{u}(t)= [\delta \hat Q,\delta \hat P,\delta \hat X,\delta \hat Y]^T $,  are given by the linearized QLEs which one can write them as  
\begin{equation}
\delta \dot{\textbf{u}}(t)=\textbf{A} \delta \textbf{u}(t)+\textbf{n}(t) 
\end{equation}
where
\begin{equation}
\textbf{A}=\left(\begin{array}{cccccc}
    -\gamma_{c} & \Omega_c &0 &0 \\
   -(\Omega_c+\omega_{sw})  & -\gamma_{c} & -\sqrt{2}G\alpha_{s}&0\\
     0 &0& -\kappa &\Delta\\
    -\sqrt{2}G\alpha_{s}& 0&-\Delta &-\kappa
\end{array}\right),
\label{drift}
\end{equation}
is the drift matrix and $\textbf{n}(t)=[\gamma_{c}(2 n_c+1),\gamma_{c}(2n_c+1),\kappa, \kappa]^T$
describes the vector of the noises. We note that the current system is stable and reaches a steady state after a transient time if all the eigenvalues of the drift matrix $ \textbf{A} $ have negative real part. These stability
conditions can be obtained by using the \textit{Routh-Hurwitz}
criterion~\cite{gry}.

The output mode of the optical cavity is given by the standard input-output relation $ \hat a_{\mathrm{out}}=\sqrt{2\kappa}\hat a-\hat a_{\mathrm{in}} $. One can also define the selected output
mode by means of the causal filter function $\hat  a^{\mathrm{filt}}=\int_{t_0}^{t}F(t-s)\hat a_{\mathrm{out}}(s) ds$, where the causal filter function
$ F(t) =\sqrt{2/\tau}\mathrm{exp}[-(1/\tau+i\Omega)t]\Theta (t)$ is characterized by central frequency $ \Omega $, bandwidth $ 1/\tau $, and  the Heaviside step function $ \Theta (t) $~\cite{Genes, Enk, Barzanjeh}. 
 In the frequency domain, the stationary covariance matrix for the quantum fluctuations of the BEC,  and the output mode of the optical cavity variables, $\textbf{u}^{\mathrm{filt}}(t)= [\hat Q, \hat P, \hat X^{\mathrm{filt}}, \hat Y^{\mathrm{filt}}]^T $, takes the form
\begin{eqnarray}
\mathbf{V}&=&\underset{t\rightarrow\infty}{\lim}\frac{1}{2}\left\langle
u_{i}^{\mathrm{filt}}(t)u_{j}^{\mathrm{filt}}(t)+u_{j}^{\mathrm{filt}}(t)u_{i}^{\mathrm{filt}}(t)\right\rangle\nonumber\\
&=& \int  d\omega\boldsymbol{\Upsilon}(\omega
)\Big(\mathbf{\tilde{M}}(\omega)+\mathbf{P}%
\Big)\nonumber\\
&&\times\mathbf{D}\Big(\mathbf{\tilde{M}}(\omega)^{\dagger
}+\mathbf{P}\Big)\boldsymbol{\Upsilon}^{\dagger}(\omega),
\label{vmat}
\end{eqnarray}
where
$\tilde{\textbf{M}}(\omega)=(i\omega \textbf{I}+\textbf{A})^{-1}$, ${\textbf{P}}=\mathrm{Diag}[0,0,\frac{1}{2\kappa},\frac{1}{2\kappa}]$ ,\\
 $\textbf{D}=\mathrm{Diag}[\gamma_c(2n_c+1), \gamma_c(2n_c+1), 2\kappa, 2\kappa]$,\\
 and $\boldsymbol{\Upsilon}(\omega)$ is the Fourier transform of
\begin{equation}
\boldsymbol{\Upsilon}(t)=\left(\begin{array}{cccccc}
   \delta(t) &0 &0 &0\\
   0&\delta(t)&0&0\\
   0 &0& \mathcal{R}&-\mathcal{I}\\
   0& 0&-\mathcal{I}& \mathcal{R}
  \end{array}\right),
\label{T}
\end{equation}
where $\mathcal{R}=\sqrt{2\kappa}\mathrm{Re}[h(t)]$ and $\mathcal{I}=\sqrt{2\kappa}\mathrm{Im}[h(t)]$ are determined by the causal filter functions. 

\section{Generating quantum discord between two distant BECs with Bell-like detection}\label{protocol}
In this section, we use the results obtained in previous section in order to create a quantum link between two remote BECs at different places by performing Bell-like detection. First we find covariance matrix of the system after Bell-like detection and then we quantify the generated bipartites quantum discord.
\subsection{Bell-like detections with arbitrary quantum efficiencies}
The overarching goal of this subsection is devoted to finding the covariance matrix of the system after Bell-like detection. More information about general aspect of this approach can be found in Ref. \cite{Stefano}. 

The whole system, as sketched in Fig.~\ref{f1}(b), is composed of two independent
bipartite bosonic modes~(include BEC, and optical field modes), where each of them possessed by Alice and Bob at different stations. Note that the bipartite bosonic modes at each site is fully characterized by covariance matrix~(\ref{vmat}). Alice and Bob prepare a bipartite state and each shares one travelling optical mode~(the outputs of optical cavities) with Charlie, who is located for simplicity halfway between them. The covariance matrix of the whole system, composed of two independent bipartite bosonic modes, can be written in the blockform
\begin{equation}\label{VoutT}
\textbf{V}_{ACB}=\left( \begin{array}{cccc}
\mathbf{A'}&\mathbf{C}\\
\mathbf{C}^T&\mathbf{B'}\\
\end{array}\right), 
\end{equation}
where the $ 4\times 4 $ matrix $\mathbf{A'}$ shows the reduced covariance matrix of the BECs modes, 
\begin{equation}\label{C}
\textbf{C}=\left( \begin{array}{cc}
\mathbf{C}_1&\mathbf{C}_2
\end{array}\right), \nonumber
\end{equation}
is a real matrix, describing the correlations between BECs and the output of optical cavities, and 
\begin{equation}
\mathbf{B'}=\left( \begin{array}{cccc}
\mathbf{B}_1&\mathbf{W}\\
\mathbf{W}^T&\mathbf{B}_{2}\\
\end{array}\right), \nonumber
\end{equation}
is the reduced covariance matrix of the two optical modes received by Charlie. For simplicity let us write down the reduced covariance matrix $\textbf{B}'$
of the optical modes by setting
\begin{eqnarray}
\mathbf{B}_1&&:=\left( \begin{array}{cccc}
\alpha_1 & \alpha_3\\
\alpha_3 & \alpha_2\\
\end{array}\right), \,\,
\mathbf{B}_2:=\left( \begin{array}{cccc}
\alpha_1' & \alpha_3'\\
\alpha_3' & \alpha_2'\\
\end{array}\right),\nonumber\\
\mathbf{W}&&:=\left( \begin{array}{cccc}
\beta_1 & \beta_3\\
\beta_3 & \beta_2\\
\end{array}\right). \nonumber
\end{eqnarray}
Note that matrices $ \textbf{A}',\, \textbf{B}', $ and $\textbf{ C}$ are determined by combination of two covariance matrices $ \textbf{V} $~(see Eq. (\ref{vmat})) of the each remote side. 

Now, Charlie applies a Bell-like detection on the received optical modes. As depicted in Fig.~\ref{f1}(b), this detection consists in
applying a beam splitter of transmissivity $ T $, which transforms the optical modes $  X_A$ and $X_B $ into the output modes
$ X_1 $ and $ X_2 $, followed by two conjugate ($I_1$ and $ I_2 $) homodyne detections. Here, we assume the realistic detectors with quantum efficiencies 
$ 0<\eta_1\leq1 $ and  $ 0<\eta_2\leq1 $. Means that two homodyne detectors modelled by inserting two beam-splitters
with transmissivities $ \eta_1 $ and $\eta_2$, which mix the
signal modes with vacuum. As a result, after Bell-like detection the final covariance matrix of the remaining modes, i.e., two BECs is given by~\cite{Stefano}
\begin{equation}\label{CMT}
\textbf{V}_{AB}=\textbf{A}'-\frac{1}{\mathrm{det}\mathbf{\Gamma}}\sum_{i,j=1}^2 \mathbf{C}_i \mathbf{K}_{ij}\mathbf{C}_j^T,
\end{equation}
where 
\begin{eqnarray}
\mathbf{K}_{11}&&:=\left( \begin{array}{cccc}
(1-T)\gamma_2 & \sqrt{T(1-T)}\gamma_3\\
\sqrt{T(1-T)}\gamma_3 &T\gamma_1\\
\end{array}\right),\nonumber\\
\mathbf{K}_{22}&&:=\left( \begin{array}{cccc}
T\gamma_2 & -\sqrt{T(1-T)}\gamma_3\\
-\sqrt{T(1-T)}\gamma_3 &(1-T)\gamma_1\\
\end{array}\right),\nonumber\\ 
\mathbf{K}_{12}&&=\mathbf{K}_{21}^T:=\left( \begin{array}{cccc}
 -\sqrt{T(1-T)}\gamma_2 &(1-T)\gamma_3\\
-T\gamma_3 & -\sqrt{T(1-T)}\gamma_1\\
\end{array}\right),\nonumber
\end{eqnarray}
with 
\begin{eqnarray}
\mathbf{\Gamma}:=\left( \begin{array}{cccc}
\gamma_1 & \gamma_3\\
\gamma_3 & \gamma_2\\
\end{array}\right),\nonumber
\end{eqnarray}
and 
\begin{eqnarray}
\gamma_1  &:=& (1-T)\alpha_1+T \alpha_1'-2\sqrt{T(1-T)}\beta_1+\frac{1-\eta_1}{\eta_1},\nonumber\\
\gamma_2  &:=&T\alpha_2+(1-T) \alpha_2'+2\sqrt{T(1-T)}\beta_2+\frac{1-\eta_2}{\eta_2},\nonumber\\
\gamma_3  &:=&\sqrt{T(1-T)}(\alpha_3'-\alpha_3)-(1-T) \beta_3+T\beta_4.\nonumber
\end{eqnarray}
 
Note that the covariance matrix~(\ref{CMT}) includes all information about the quantum correlations between two BECs after Bell-like detection. The rest of the paper devotes mainly to study these quantum correlations and we explicitly show that the Bell-like detection ables to  generate both quantum discord and quantum entanglement between distant BECs. 
\subsection{quantum discord between two BECs after Bell-like detection}
 In this subsection, we focus on quantum discord and we show that how Bell-like detection is able to create the quantum discord between two remote BECs. 

The quantum discord between BEC $ A $ and BEC $ B $, is defined as~\cite{Zurek, Zurek2} 
\begin{eqnarray}\label{dis}
\mathcal{D}=\mathcal{I}(AB)-\mathcal{J}(B|A).
\end{eqnarray}
where $\mathcal{I}({AB})$ represents the quantum correlation between BEC $ A $ and BEC $ B $ and $\mathcal{J}(A|B)$ is interpreted as the information gain about BEC $ B $ by measuring the BEC $  A$. According to~\cite{lloyd} on the optimality of Gaussian discord, the
quantum discord Eq.~(\ref{dis}) can be computed in the term of elements of covariance matrix~(\ref{CMT}), reads~\cite{adesso,Paris,Farace}
\begin{eqnarray}\label{discord}
\mathcal{D}=f(\sqrt{s_1})-f(\lambda_{-})-f(\lambda_{+})+f(\sqrt{\epsilon}),
\end{eqnarray}
where
\begin{eqnarray}
f(x)&=&\left( \frac{x+1}{2}\right)\log_2\left( \frac{x+1}{2}\right) -\left( \frac{x-1}{2}\right)\log_2\left( \frac{x-1}{2}\right),\nonumber\\
\end{eqnarray}
and if $  (s_4-s_2s_1)^2\leq (1+s_1)s_3^2(s_2+s_4) $ then
\[\epsilon=\frac{2s_3^2+(s_1-1)(s_4-s_2)+2|s_3|\sqrt{s_3^2+(s_1-1)(s_4-s_2)}}{(s_1-1)^2}, \]
otherwise 
\[\epsilon=\frac{s_2s_1-s_3^2+s_4-\sqrt{s_3^4+(s_4-s_2s_1)^2-2s_3^2(s_4+s_2s_1)}}{2s_1},\]
where $ s_1=4\,\mathrm{det}\textbf{V}_1,\, s_2=4\,\mathrm{det}\textbf{V}_2,\, s_3=4\,\mathrm{det}\textbf{V}_3,\, s_4=16\,\mathrm{det}[\textbf{V}_{AB}]$. Here we have rewritten the two-modes covariance matrix~(\ref{CMT}) as
\begin{equation}\label{CMT2}
\textbf{V}_{AB}=\left( \begin{array}{cccc}
\textbf{V}_1&\textbf{V}_3\\
\textbf{V}_3^T&\textbf{V}_2
\end{array}\right), 
\end{equation}
where $\lambda_{\pm}=2^{-1/2}\left[ {s_{\Delta}\pm\sqrt{s_{\Delta}^2-4s_4}}\right] ^{1/2}$ and $s_{\Delta}=s_1+s_2+2s_3$.

 \subsection{Results}\label{Quantum correlation}
\begin{figure}[ht]
\centering
\includegraphics[width=3.5in]{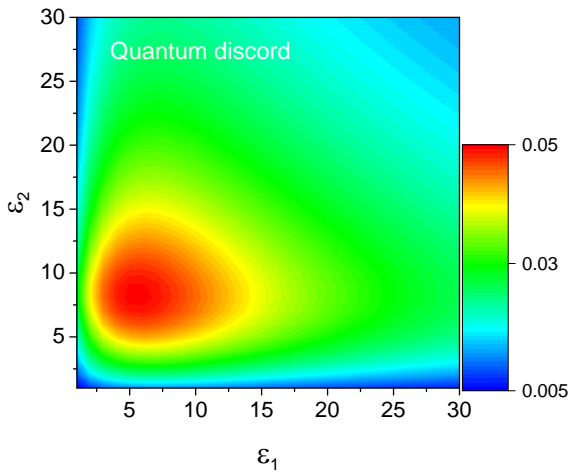}
\caption{(Color online)~Quantum discord between remote BECs after Bell-like detection with respect to the filtering bandwidths $\varepsilon_1=\Omega_1 \tau_1 $ and $\varepsilon_2=\Omega_2 \tau_2 $, where $ \eta_1=\eta_2\simeq1 $. The central frequencies of detected output fields are $\Omega_1=\Omega_2=-\omega_B$. Here we assume an optical cavity of length $L=1$ mm with wavelength of $\lambda=1046$ nm, finesse $\mathcal{F}=1.15\times 10^5$, and damping rate $\kappa=\pi c/L\mathcal{F}$. The optical cavity is driven with pump amplitude $E_\mathrm{d}=3\kappa$ and its detuning is $\Delta=\Omega_c$. The recoil frequency of BEC is $\omega_R/2\pi=3.57$ kHz with dissipation of collective density excitations $\gamma_{c}=0.001\kappa$, atom-cavity coupling constant $G=0.5\omega_B$ and temperature $T_{c}=0.1 \mu$K. }
\label{figDiscord}
\end{figure}

We now utilize the results obtained in the previous subsection to show that how Bell-like detection can generate quantum discord between two distant sides. The bipartite quantum discord between subparts of the remote systems, i.e., BEC-BEC is determined by Eq.~(\ref{discord}). The quantum discord $\mathcal{D}$ after Bell-like detection as a function of the filtering bandwidths $\varepsilon_1=\Omega_1 \tau_1 $ and $\varepsilon_2=\Omega_2 \tau_2 $ has been plotted in Fig.~\ref{figDiscord}, where we have
taken the experimentally feasible parameters~\cite{Bren, Szirmai} i.e., an optical cavity of length $L=1$ mm with wavelength of $\lambda=1046$ nm, finesse $\mathcal{F}=1.15\times 10^5$, and damping rate $\kappa=\pi c/L\mathcal{F}$. The optical cavity is driven with pump amplitude $E_\mathrm{d}=3\kappa$ and its detuning is $\Delta=\Omega_c$. The recoil frequency of BEC is $\omega_R/2\pi=3.57$ kHz with dissipation of collective density excitations $\gamma=0.001\kappa$, atom-cavity coupling constant $G=0.5\omega_B$, and temperature $T_c=0.1 \mu$K. As seen, the Bell-like detection successfully generates the quantum discord among the elements of two distant BECs. Fig.~\ref{figDiscord}, however,  shows that the quantum discord of BEC-BEC is considerable around $\varepsilon_1=\varepsilon_2\simeq 8$, which means that, in practice, by appropriately filtering the output fields one can generate stronger quantum discord with Bell-like detection. 

\begin{figure}[ht]
\centering
\includegraphics[width=2.8in]{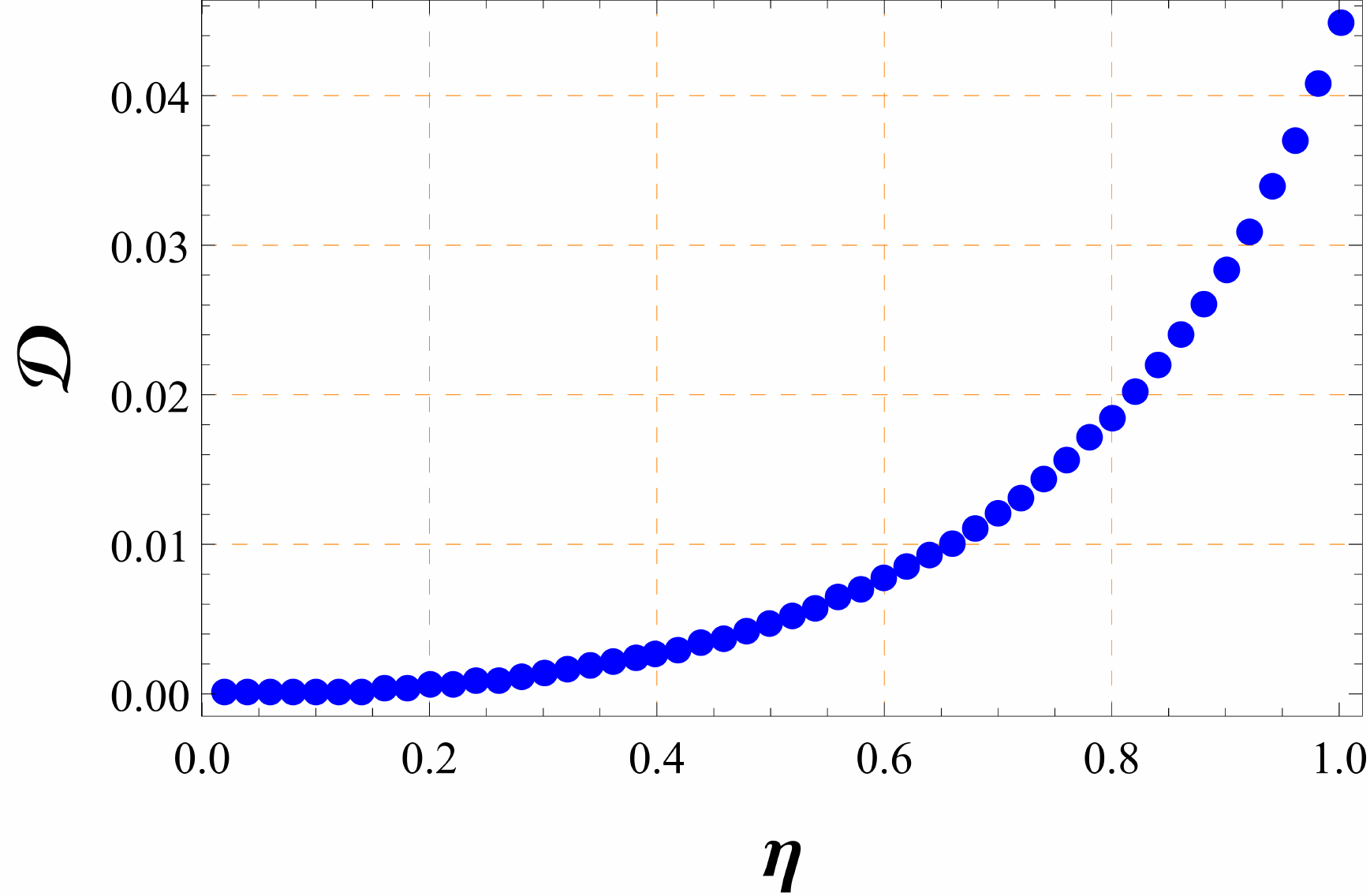}
\caption{(Color online) Effect of the detectors quantum efficiencies $ \eta:=\eta_1=\eta_2 $ on the quantum discord  between distant BECs where we assume $\varepsilon_1=\varepsilon_2=8$. The other parameters are the same as Fig.~\ref{figDiscord}.}
\label{efficency}
\end{figure}

As is noted,  we consider non-ideal Bell-like detection in which we assume the realistic homodyne detectors with quantum efficiencies 
$ \eta_1<1 $ and  $ \eta_2<1 $. The effect of the quantum efficiencies of the detectors on the quantum discord between distant nodes is plotted in Fig.~\ref{efficency} where we assume $ \eta:=\eta_1=\eta_2 $. This figure confirms that in order to establish strong quantum discord between two remote BECs one should use close to perfect detectors with quantum efficiencies $ \eta_1\simeq1 $ and  $ \eta_2\simeq1 $. 

\begin{figure}[ht]
\centering
\includegraphics[width=3.5in]{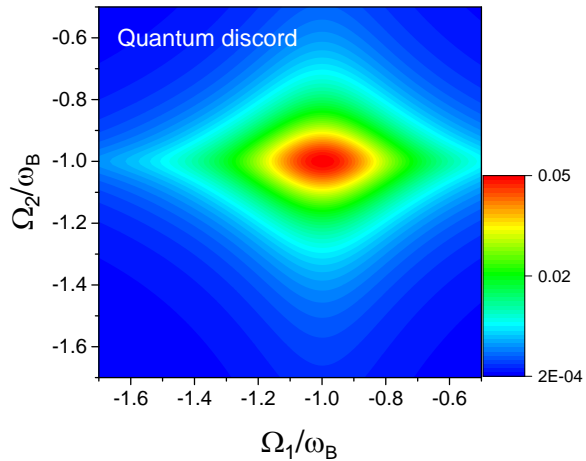}
\caption{(Color online) Quantum discord between two remote BECs with respect to the normalized frequencies $\Omega_1/\omega_B $ and $\Omega_2/\omega_B$, where we assume $ \eta_1=\eta_2\simeq1 $ and $\varepsilon_1=\varepsilon_2= 8$. The other parameters are the same as Fig.~\ref{figDiscord}. }
\label{figdiscord}
\end{figure}

\begin{figure}[ht]
\centering
\includegraphics[width=2.8in]{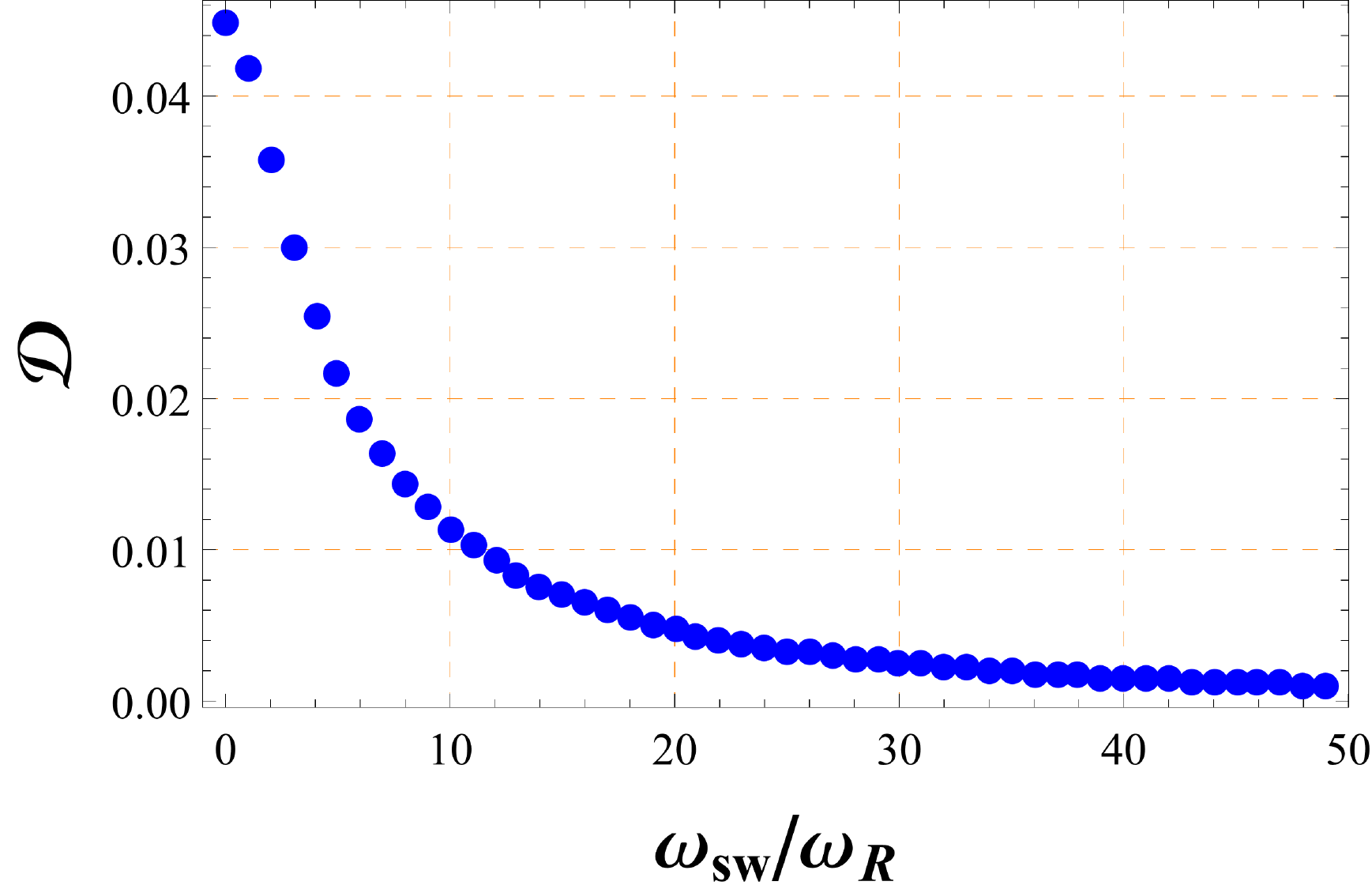}
\caption{(Color online) Quantum discord between remote BECs with respect to the collision parameter $ \omega_{sw}/\omega_R $ for $\varepsilon_1=\varepsilon_2=8$. The other parameters are the same as Fig.~\ref{figDiscord}.}
\label{Dcol}
\end{figure}

Moreover, Fig.~\ref{figdiscord} shows the quantum discord between two BECs with respect to $ \Omega_1/\omega_B $ and  $ \Omega_2/\omega_B $ for $\omega_{sw}=0$, where the discord is maximum around the particular resonance frequency of BECs mode $ \Omega_i=-\omega_B $~(with $ i=1,2 $). Finally, the effect of atomic collisions has been plotted in Fig.~\ref{Dcol}. It is clear that the effect of the atomic collisions on the quantum discord is inevitable. The atom-atom interactions in the BEC strongly degrades the bipartite quantum discord between remote nodes.  
\section{Entanglement between two distant BECs}\label{entanglement}
Finally, we focus on the the entanglement after Bell-like detection. The bipartite entanglement between two remote BECs after Bell-like detection can be characterized by the logarithmic negativity~\cite{eisert}
\begin{eqnarray}\label{EN}
{E_{\mathcal{N}}}=\max[0,-\ln(2\eta_{-})],
\end{eqnarray}
where $\eta_{-}=2^{-1/2}\left[ {\sigma-\sqrt{\sigma^2-4 \mathrm{det}\textbf{V}_{AB}}}\right] ^{1/2}$ is the least symplectic eigenvalue of the covariance matrix~(\ref{CMT2}) and $\sigma=\mathrm{det}\textbf{V}_1+\mathrm{det}\textbf{V}_2-2\mathrm{det}\textbf{V}_3$.
\begin{figure}[ht]
\centering
\includegraphics[width=3.5in]{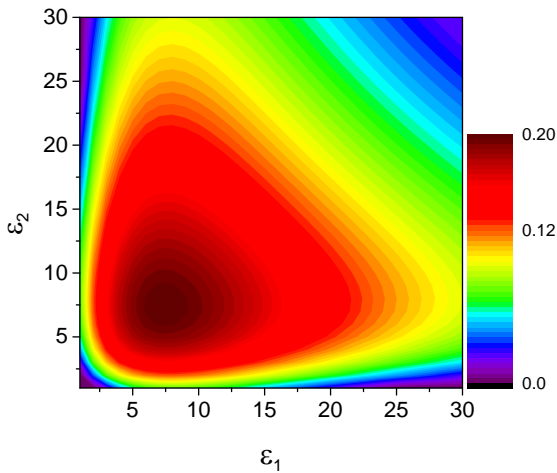}
\caption{(Color online) Bipartite entanglement after Bell-like detection between remote BECs with respect to the filtering bandwidths $\varepsilon_1=\Omega_1 \tau_1 $ and $\varepsilon_2=\Omega_2 \tau_2 $, where $ \eta_1=\eta_2\simeq1$. The central frequencies of detected output fields are $\Omega_1=\Omega_2=-\omega_B$. The other parameters are the same as Fig.~\ref{figDiscord}.}
\label{figentang1}
\end{figure}

\begin{figure}[ht]
\centering
\includegraphics[width=2.8in]{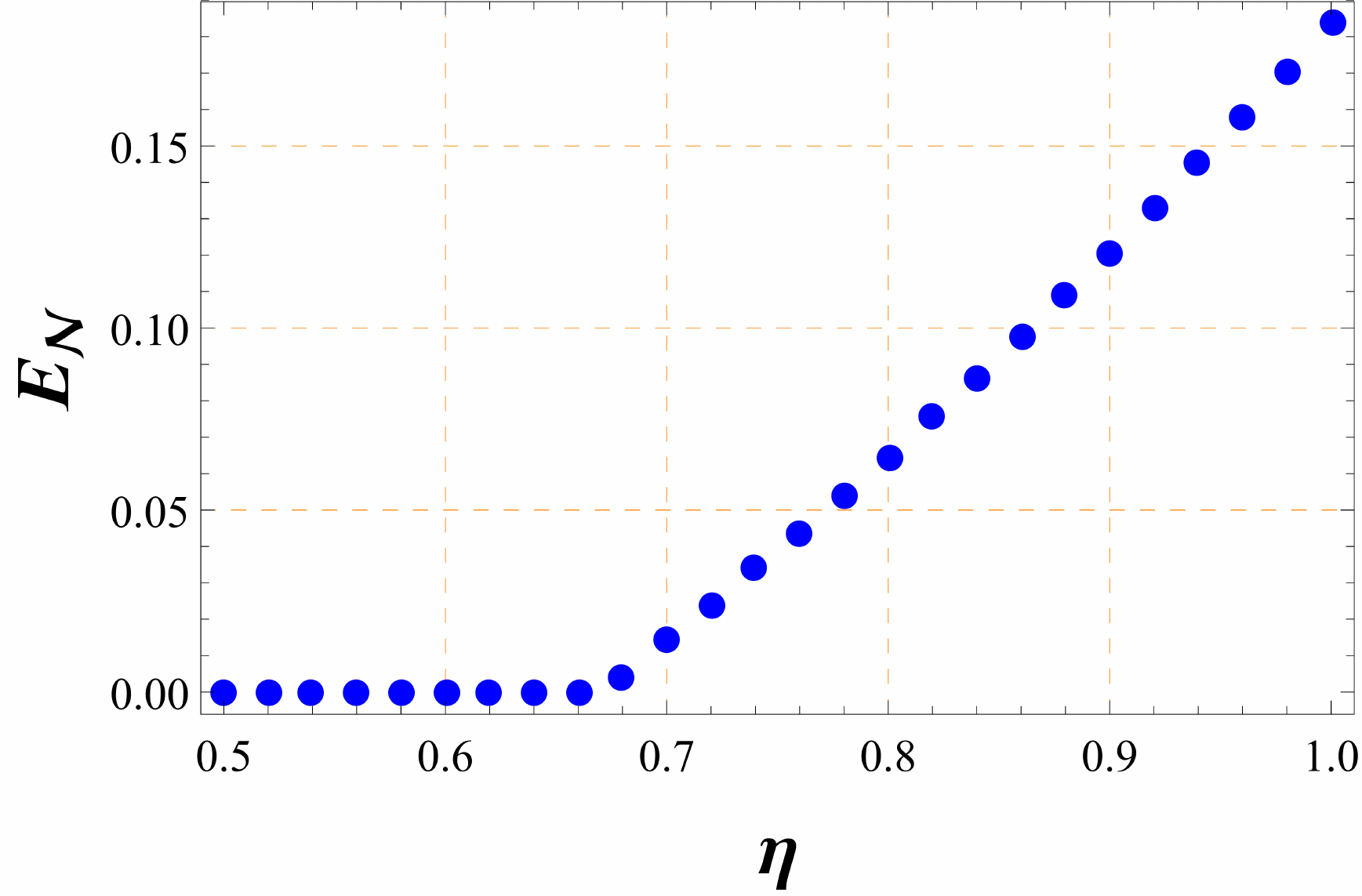}
\caption{(Color online)~Effect of the detectors quantum efficiencies $ \eta:=\eta_1=\eta_2 $ on the generated entanglement between distant two BECs where we assume $\varepsilon_1=\varepsilon_2=8$. The other parameters are the same as Fig.~\ref{figDiscord}.}
\label{pefficency}
\end{figure}

 The logarithmic negativity $E_\mathcal{N}$ after Bell-like detection as a function of the filtering bandwidths $\varepsilon_1=\Omega_1 \tau_1 $ and $\varepsilon_2=\Omega_2 \tau_2 $ has been plotted in Fig.~\ref{figentang1}. As seen, the Bell-like detection successfully generates entanglement between distant BECs. However, as we excepted this figure shows that the entanglement between two remote BECs is considerable around $\varepsilon_1=\varepsilon_2\simeq8$. Furthermore, the effect of the quantum efficiencies of the detectors on the entanglement between distant nodes is plotted in Fig.~\ref{pefficency}. 
 
\section{conclusions}\label{conclusion}
In this paper, we have studied the possibility of producing the quantum correlations between two distant Bose-Einstein condensates simply by performing a non-ideal Bell-like detection. We have shown that by applying a Bell-like detection on the travelling optical fields of two optical cavities each containing a Bose-Einstein condensate, one can generate quantum discord between distant Bose-Einstein condensates. The covariance matrix of the Bose-Einstein condensate modes after Bell-like detection has been found. We then have evaluated the quantum discord between two distant Bose-Einstein condensates at different nodes. The effect of the filtering bandwidths and quantum efficiencies of the detectors have been discussed. Moreover, the effect of the atom-atom interaction in the Bose-Einstein condensates on the quantum discord has been investigated. 
We have also studied the entanglement in the system that is created by the Bell-like detection. The effect of the detection bandwidths and  efficiencies of detectors on the entanglement have been studied. The promising results of the paper show that the Bell-like detection is able to build up the quantum correlations~(discord and entanglement) between distant BECs that never interacted. However, we should note that the degree of these correlations that can be achieved with this technique still is not very large. 

In view of significant recent progress in quantum state
engineering, the Bell-like detection protocol discussed here may offer
interesting perspectives for future applications in optical implementations
of hybrid quantum repeaters and quantum communication
networks. Finally, we note that the Bell-like detection scheme discussed here can be implemented experimentally by employing the standard protocols of entanglement swapping~\cite{Bou, Zukowski, Zei}. 
\section*{Acknowledgements}
 The authors would like to thank D. P. DiVincenzo, S. Pirandola, A. Mari and A. Farace for valuable discussions.
M.E.A and M.A.S wish to thank the Office of Graduate
Studies of The University of Isfahan for their support. The work of S.B has been supported by the Alexander von Humboldt
foundation. 
\bibliography{BEC}
\bibliographystyle{apsrev4-1}

\end{document}